\documentclass[11pt]{article}
\usepackage[dvips]{graphicx,color}
\usepackage{amssymb}
\usepackage{amsfonts}
\usepackage{latexsym}
\usepackage{amsmath}
\graphicspath{{/}}
\begin{document}
\begin{center}
{\Large{\bf{ Nonequilibrium Approach to Bloch-Peierls-Berry Dynamics } }} \\

\vspace{0.3in} John C. Olson and Ping Ao\footnote{Corresponding Author. E-mail: aoping@u.washington.edu; ph: (206) 543-7837} \hspace{0.5cm}
\\

Mechanical Engineering Department\\
University of Washington \\
Seattle, Washington 98195, USA \\

May 08, 2006
\end{center}


We examine the Bloch-Peierls-Berry dynamics under a classical
nonequilibrium dynamical formulation. In this formulation all
coordinates in phase space formed by the position and crystal
momentum space are treated on equal footing. Explicitly
demonstrations of the no (naive) Liouville theorem and of the
validity of Darboux theorem are given. The explicit equilibrium
distribution function is obtained. The similarities and
differences to previous approaches are discussed. Our results
confirm the richness of the Bloch-Peierls-Berry dynamics.



\section{Introduction}

One of the fundamental dynamical equations in condensed matter
physics is the so-called Bloch equation, describing one electron
moving in a periodic potential in position and crystal momentum
space \cite{ashcroft}:
$ \dot{\bf r}
 = {1\over \hbar}{\nabla_{\bf k} \phantom{i} \epsilon({\bf k})}
 \, ,
  \hbar \dot{\bf k}
   =  - e \nabla_{\bf r} \phantom{i} \phi({\bf r}) \, .
$
Here ${\bf r} \in \Re^3$ is position, ${\bf k}\in \Re^3$ crystal
momentum, $\epsilon({\bf k})\in \Re^1$ kinetic energy, $\phi({\bf
r})\in\Re^1$ the electric potential, $e$ the electric charge of an
electron, and $\hbar$ the Planck constant. The index $l$ in the
gradient operation, $\nabla_l \equiv { \partial / \partial l}$
indicates the corresponding space coordinate.
A scalar function $\mathcal{H}({\bf r},{\bf k})$, termed the Bloch
Hamiltonian, can be defined as
\begin{equation} \label{hamiltonian}
 \mathcal{H}({\bf r}, {\bf k}) =
   \frac{1}{\hbar} [ \epsilon({\bf k})  + e \phi({\bf r}) ] \; .
\end{equation}
In this case, the Bloch dynamics can be rewritten in the canonical
Hamiltonian form: $\dot{\bf r} = \nabla_{\bf k}
\mathcal{H}$ and $ \dot{\bf k} = - \nabla_{\bf r} \mathcal{H} $.
The usual incompressible condition, the Liouville theorem, in the
six dimensional $ \bf{x}^{\tau} \equiv ( \bf{r}, \bf{k})$ phase
space follows,
\begin{eqnarray} \label{E:H3}
  \nabla_{\bf x} \cdot \dot{\bf x}
  & = & {\nabla_{{\bf r}} \cdot \dot{\bf r}}
      + {\nabla_{\bf k} \cdot \dot{\bf k}}           \nonumber \\
  & = &  \nabla_{\bf r} \cdot \nabla_{\bf k}  \mathcal{H}
       - \nabla_{\bf k} \cdot \nabla_{\bf r}  \mathcal{H} \nonumber \\
  & = & 0
\end{eqnarray}

R. Peierls successfully extended the Bloch equations to the
case with a weak magnetic field ${\bf B}({\bf r})$
\cite{ashcroft}:
$ \dot{\bf r} = {1 \over \hbar} {\nabla_{\bf k} \phantom{i}
 \epsilon({\bf k})}$, and
$ \hbar \dot{\bf k} = - e\nabla_{\bf r} \phantom{i} \phi({\bf r})
 - e \dot{\bf r} \times {\bf B}({\bf r}) $.
The Hamiltonian in the form of Eq.(\ref{hamiltonian}) remains the
same for this extended dynamics, and the Liouville theorem of
Eq.(\ref{E:H3}) still holds. Everything appears as expected.

However, the Bloch equations of electron dynamics have been
recently modified to include both magnetic field ${\bf B}({\bf
r})$ and Berry curvature ${\bf \Omega} ({\bf k})$ in crystal
momentum space \cite{sundaram}. The following extended dynamical
equations in the semiclassical limit were obtained, termed
Bloch-Peierls-Berry equations in the present paper:
\begin{subequations}\label{E:E1}
 \begin{align}
  \dot{\bf r}  & =  {1 \over \hbar}{\nabla_{\bf k}\phantom{i} \epsilon({\bf k}) }
          - \dot{\bf  k} \times {\bf \Omega} ({\bf k})  \\
  \hbar\dot{\bf k} & =  - e\nabla_{\bf r}\phantom{i} \phi({\bf r})
          - e \dot{\bf  r} \times {\bf B}({\bf r})
 \end{align}
\end{subequations}
Interesting, unexpected, and rich
behaviors occur in the context of this novel dynamics
\cite{sundaram,xiao}. For example, it has been found that in six
dimensional phase space the current flow becomes compressible,
that is, the Liouville theorem does not hold \cite{xiao}:
\begin{equation} \label{E:H3-3}
  \nabla_{\bf{x}} \cdot \dot{\bf{x}} \neq  0 .
\end{equation}
Thus the straight-forward conventional means to Hamiltonian
analysis does not appear available. Such a situation has been called
non-Hamiltonian dynamics \cite{xiao}.

To remedy this problem of non-Hamiltonian dynamics, Xiao et al.
\cite{xiao} introduced a density correction factor, denoted
$J({\bf r}, {\bf k})= 1 + {e \over \hbar} {\bf B}\cdot{\bf \Omega}
$ in the present paper, to force the divergence in phase space to
zero:
\begin{equation}
 \nabla_{\bf{r}} \cdot (J({\bf r}, {\bf k}) \, \dot{\bf{r}})
   + \nabla_{\bf{k}} \cdot (J({\bf r}, {\bf k}) \, \dot{\bf{k}}) = 0
\end{equation}
Of primary motivation using this technique was in the definition
of a phase distribution function consistent with the Liouville
theorem, because the Liouville theorem states that phase volume in
a canonical Hamiltonian formulation is conserved along it's
trajectory.  This has some good properties. For example, with
the Liouville theorem a distribution function over a system phase,
$\bf{x}$, will be time-invariant.  To verify this property we
begin with the total time derivative of a distribution $f({\bf x}
, t)$ in phase space,
$  {df\over dt} = {\partial f \over \partial t} + \dot {\bf x}
\cdot \nabla_{\bf x}f $
and the continuity equation,
$   {\partial f \over \partial t} + \nabla_{\bf x}\cdot(\dot {\bf
x} f) = 0$.
When the Liouville theorem holds, $ \nabla_{\bf x} \cdot \dot{\bf x} = 0
$, the continuity equation is identical to the
time-invariant distribution, $ {df / dt} = 0$.

Disagreement on the treatment in Ref.\cite{xiao} exist in
literature \cite{bliokh,duval,ghosh,gosselin}. For example, in a
comment from Duval et al., \cite{duval}, an alternate approach was
pointed out such that a local canonical form, and likewise a
Hamiltonian description, is achievable. In addition to the
question of mathematical formulation, there is a real issue of
possibly different physical consequences in different approaches
\cite{xiaoreply}.

In the present paper we study the problem from a completely
different point of view to see how the Hamiltonian like structure
and the equilibrium distribution function emerge from a
nonequilibirum process: the Darwinian dynamics \cite{ao2005}. We
will show that a Hamiltonian, or energy function, naturally emerges.
In our demonstration it is clear that the Berry phases due to the
magnetic field and the Berry curvature of the momentum
space appears in equal footing. Furthermore, such a procedure
provides a straightforward discussion on the equilibrium
distribution: a nonequilibrium setting provides a natural way to
select the steady state distribution. Not all of our results
are new. Nevertheless, our demonstration appears to provide a
clear, consistent, but completely classical starting point to
detail the system characteristics as previously described while
possessing no proclivity toward an incompressible or canonical
phase.  For example, a direct solution to the Fokker-Planck
equation has been detailed.  This solution is found independent of
the Liouville theorem and may offer a unique insight into a
probability distribution.

We begin the remainder of the paper in section 2 by first
describing the generic decomposition of Darwinian dynamics and
arranging the Berry modified Bloch equations, Eq.~(\ref{E:E1}), to
this form. In section 3 we will present the divergence
analytically to show that the system is compressible. In section 4
we evaluate the Jacobi identity to show that a local canonical
form exists and in section 5 we reveal the evolution of a
probability distribution developed directly from the general form
of Darwinian dynamics.  Section 6 is a summary and discussion.

\section{Evolutionary Decomposition and Conservation of \lq\lq Energy" }

\subsection{Darwinian dynamics}

The Darwinian dynamics arises from a generic nonequilibrium
process common in biological, physical, and social sciences
\cite{ao2005}. One of it's main features is to treat all dynamical
variables on an equal footing. This is typically achieved through
expression of the dynamics in a set of first order
stochastic differential equations.

Given a generic first order dynamic system of states, $\bf x$,
separated into the deterministic, ${\bf f}({\bf x})$, and
stochastic, ${\bf \zeta}({\bf x}, t)$, components:
\begin{equation} \label{E:Darwin0}
    \dot{\bf x} = {\bf f}({\bf x}) + {\bf \zeta}({\bf x}, t)
\end{equation}
The noise is typically approximated as Gaussian and white
with zero mean, $\langle {\bf \zeta}({\bf x}, t) \rangle = 0
$, and variance as
\begin{equation}
 \langle {\bf \zeta}({\bf x}, t){\bf \zeta}^{\tau} ({\bf x}, t') \rangle
  = 2 \; \omega \; D({\bf x}) \;  \delta(t-t') \, .
\end{equation}
Here the nonnegative constant $\omega$ plays the role of
temperature in physics. Further characterization of the noise
comes from the diffusion matrix $D$.

There then exists a unique decomposition as follows
\cite{ao2005,ao2004,kat,ya2006}:
\begin{equation} \label{E:Darwin}
 [S({\bf x}) + T({\bf x})] \dot{\bf x}
  = - \nabla_{\bf x} \Phi({\bf x}) + {\bf \xi}({\bf x}, t)
\end{equation}
Here $S$ is a symmetric and positive definite friction matrix,
related to the zero-mean Gaussian and white noise as follows
\begin{equation}
 \langle {\bf \xi}({\bf x}, t){\bf \xi}^{\tau} ({\bf x}, t')
  \rangle = 2 \; \omega \; D({\bf x}) \;  \delta(t-t') \, .
\end{equation}
$T({\bf x})$ is an antisymmetric matrix, $\Phi({\bf x})$ a scalar
potential function and $S$ a symmetric matrix. By definition $S$
is positive semi-definite, $\dot{\bf  x}^{\tau} S \dot {\bf x} \!
\geq\!0$, and will have a dissipative effect on the potential
function, showing a tendency to approach the potential minima. The
$T$ matrix describes a non-dissipative part of the dynamics, $\dot
{\bf x}^{\tau}T \dot {\bf x}$ = 0, and tends to conserve the
potential function.

\subsection{ Reformulation of Bloch-Peierls-Berry equations}

Now we reformulate the Bloch-Peierls-Berry equations from the point
of view of Darwinian dynamics, treating position and momentum
coordinates in equal footing. We begin the decomposition of these
equations~(\ref{E:E1}) by arranging
them in terms of the complete state vector $\bf x$ in the form of
Darwinian dynamics, Eq.(\ref{E:Darwin}). The friction matrix
$S$ may not be known. We assume its existence and may take it to
be zero at the end of calculation whenever needed. The existence
of electron-phonon interaction and other dissipative processes and
their small possible effect in a solid justifies such a procedure.
Beginning from this formation of our descriptive matrices a potential
function will then become apparent.

We rewrite the Bloch-Peierls-Berry equations, Eq.(\ref{E:E1}):
\begin{eqnarray} \label{E:MAT1}
 \dot{\bf x}
  & \equiv & \begin{pmatrix}\dot{\bf r} \\
              \dot{\bf  k} \end{pmatrix} \nonumber \\
  & = & \begin{pmatrix}
       {1 \over \hbar}{\nabla_{\bf k} \phantom{i} \epsilon({\bf k})} \\
         - {e \over \hbar}\nabla_{\bf r} \phantom{i} \phi({\bf r})
        \end{pmatrix}
      - M({\bf x}) \dot{\bf x}
\end{eqnarray}
Here the matrix $M \in \Re^{6\times 6}$ contains the effects of
both magnetic field ${\bf B}$ in position space and Berry
curvature ${\bf \Omega}$ in momentum space.
\begin{equation}
 { M({\bf x}) =
  \begin{pmatrix}0 & 0 & 0 & 0 & \Omega_{3} & -\Omega_{2}
    \\ 0 & 0 & 0 & -\Omega_{3} & 0 & \Omega_{1}
    \\ 0 & 0 & 0 &  \Omega_{2} &  -\Omega_{1} & 0
    \\ 0 & {e\over \hbar}B_{3} & -{e\over \hbar}B_{2} & 0 & 0 & 0
    \\ -{e\over \hbar}B_{3} & 0 & {e\over \hbar}B_{1} & 0 & 0 & 0
    \\ {e\over \hbar}B_{2} & -{e\over \hbar}B_{1}& 0 & 0 & 0 & 0
  \end{pmatrix}}
\end{equation}
With $I$ as an identity matrix, Eq.(\ref{E:MAT1}) becomes:
\begin{equation} \label{E:MAT2}
    {(I + M({\bf x})) \dot {\bf x} = \begin{pmatrix}
              {{1 \over \hbar}\nabla_{\bf k}\phantom{i} \epsilon(\bf k)} \\
              - {e \over \hbar}\nabla_{\bf r}\phantom{i} \phi({\bf r})
                      \end{pmatrix}}
\end{equation}
This is almost in the form of Eq.(\ref{E:Darwin}) of Darwinian
dynamics and the right hand side appears
similar to the gradient of the Bloch Hamiltonian $\mathcal{H}$, Eq.~(\ref{hamiltonian}). Suggesting
that the Hamiltonian or energy function in the Bloch-Peierls-Berry
dynamics may indeed be the original Hamiltonian of
Eqs.~(\ref{E:E1}). This makes a notable potential function because
it is a straightforward representation of the total energy in the
system, we neglect the constants $e$ and $\hbar$ when speaking of
total energy because they can simply be absorbed into $\bf x$.

We also must mind the order and sign of the state vector $\bf x$
compared to the right side of Eq.~(\ref{E:MAT2}). To be consistent
an orthogonal transformation matrix $R$ is applied to both sides of
the equation (\ref{E:MAT2}).
\begin{equation}
 {R = \begin{pmatrix}
    0 & 0 & 0 & -1 & 0 & 0 \\
    0 & 0 & 0 & 0 & -1 & 0 \\
    0 & 0 & 0 & 0 & 0 & -1 \\
    1 & 0 & 0 & 0 & 0 & 0 \\
    0 & 1 & 0 & 0 & 0 & 0 \\
    0 & 0 & 1 & 0 & 0 & 0
       \end{pmatrix}}
\end{equation}
The right side of Eq.(\ref{E:MAT2}) becomes
\begin{equation}\label{E:Com}
\begin{split}
 R {\begin{pmatrix} {{1\over \hbar}\nabla_{\bf k}\phantom{i} \epsilon(\bf k)}
    \\ -{e\over\hbar} \nabla_{\bf r}\phantom{i} \phi(\bf r)\end{pmatrix}}
    & =
    {\begin{pmatrix}{e\over\hbar} \nabla_{\bf r}\phantom{i} \phi(\bf r) \\
     {{1\over \hbar}\nabla_{\bf k}\phantom{i} \epsilon(\bf k)}\end{pmatrix}} \\
    & = \nabla_{\bf x}\mathcal{H}(\bf x)
\end{split}
\end{equation}
and the left side contains an antisymmetric matrix $T$:
\begin{equation}
 R [I + M({\bf x})] \dot {\bf x} =
  T({\bf x}) \dot {\bf x}
\end{equation}
where
\begin{equation}
 {T({\bf x}) = \begin{pmatrix}
     0 & -{e\over \hbar}B_{3} & {e\over \hbar}B_{2} & -1 & 0 & 0
    \\ {e\over \hbar}B_{3} & 0 &-{e\over \hbar}B_{1} & 0 & -1 & 0
    \\ -{e\over \hbar}B_{2} & {e\over \hbar}B_{1}& 0 & 0 & 0 & -1

    \\ 1 & 0 & 0 & 0 & \Omega_{3} & -\Omega_{2}
    \\ 0 & 1 & 0 & -\Omega_{3} & 0 & \Omega_{1}
    \\ 0 & 0 & 1 &  \Omega_{2} &  -\Omega_{1} & 0
    \end{pmatrix}}
\end{equation}
Collecting all terms the Bloch-Peierls-Berry equation, Eq.(\ref{E:E1}), is
then transformed into the form of Darwinian dynamics,
Eq.(\ref{E:Darwin}):
\begin{equation} \label{E:dx}
 T({\bf x}) \; \dot{\bf x} = - \nabla_{\bf x} \mathcal{H}({\bf x})
\end{equation}
The $S$ matrix, as well as the diffusion matrix $D$,  is null in
this case, which may be thought as the zero electron-phonon
interaction limit. The equivalent form to Eq.(\ref{E:Darwin0}) of
the Darwinian dynamics is
\begin{equation}  \label{E:dx0}
  \dot{\bf  x} = - Q({\bf x}) \nabla_{\bf x} \mathcal{H}({\bf x}) \; ,
\end{equation}
with $Q({\bf x}) = T^{-1} ({\bf x})$.  A Poisson bracket can be
easily defined as
\begin{equation}\label{poisson}
 [f({\bf x}), g({\bf x})] \equiv
  \sum_{i,j=1}^{6} Q_{ij}({\bf x})
   \frac{\partial f}{\partial x_i} \frac{\partial g}{\partial x_j} \; .
\end{equation}
Here $f$ and $g$ are two arbitrary functions of the phase space ${\bf
x}$. The Poisson bracket is anti-symmetric because $Q$ is. With this 
Poisson bracket, the Bloch-Peierls-Berry equations are now
\begin{equation}\label{canonical}
  \dot{\bf x} = [\mathcal{H}({\bf x}), {\bf x}] \; ,
\end{equation}
the familiar form in Hamiltonian dynamics.

\subsection{Conservation of energy}

The potential function of Bloch-Peierls-Berry dynamics in the form
of Darwinian dynamics may be clearly identified as the Bloch
Hamiltonian:
\begin{equation}
  \Phi({\bf x}) = \mathcal{H}({\bf x}) \
    =  \frac{1}{\hbar} [ \epsilon({\bf k})  + e \phi({\bf r}) ] \; .
\end{equation}
It is straightforward to verify that such an \lq\lq energy" given
by the Bloch Hamiltonian is conserved in the Bloch-Peierls-Berry
dynamics, as expected:
\begin{equation}\label{E:CONS}
\begin{split}
 {d \over dt}\mathcal{H}({\bf x})
  & = {\partial {\bf x}\over \partial t} {\partial \mathcal{H}({\bf x}) \over \partial {\bf x}} \\
  & = \dot {\bf x}^{\tau}\nabla_{\bf x} \mathcal{H}({\bf x})\\
  & = \dot {\bf x}^{\tau} T({\bf x}) \dot {\bf x} \\
  & = 0
\end{split}
\end{equation}
Thus, we may indeed identify the Bloch Hamiltonian given in Eq.(2)
as the candidate for the Hamiltonian for the Bloch-Peierls-Berry
dynamics of Eq.(\ref{E:E1}), upon the clarification of two
questions raised in the literature, the validity of the Liouville
theorem, and the validity of the Darboux theorem, to be discussed
in the next two sections.

\section{Compressibility and No Liouville Theorem}

Here we will determine an analytical result for the divergence of
the Bloch-Peierls-Berry equations (\ref{E:E1}) from the Darwinian dynamics
given by Eq.~(\ref{E:dx}), and show that it is indeed non-zero and
thus compressible. This compressibility feature is pronounced and 
surprising, as pointed out in Ref.\cite{xiao}. An explicit demonstration is given
in light of the present Darwinian dynamics formulation.

The equation in question is, following Eq.(\ref{E:Darwin}):
\begin{equation} \label{E:comp}
 {\nabla_{\bf x} \cdot \dot {\bf x}
   = - \nabla_{\bf x} \cdot (T^{-1}({\bf x}) \nabla_{\bf x} \mathcal {H}({\bf x}))}
\end{equation}
We first note that because $T^{-1}$ is antisymmetric only the
divergence of the $T^{-1}$ matrix on the right hand side of the
equation need be considered:
\begin{equation}\label{E:MAT4}
 {\nabla_{\bf x} \cdot \dot {\bf x}
  = (\nabla_{\bf x} \cdot T^{-1})\nabla_{\bf x} \mathcal{H}({\bf x})}
\end{equation}

Next, we note a useful identity expressing the divergence of
$T^{-1}$ as a function of the divergence of $T$, because $T$ is
easier to operate on.  This identity is found as follows,
where $I$ is the identity matrix:
\begin{equation} \label{E:IS}
 \begin{split}
  \nabla \cdot I \quad
   & =  \nabla_{\bf x} \cdot (T^{-1}T) \\
   & = (\nabla_{\bf x} \cdot T^{-1})T
      + \sum_{i = 1}^{6} T^{-1}(i,:)
       { \partial T\over \partial x_{i} } \\
   & = 0
\end{split}
\end{equation}
The $T^{-1}(i,:)$ denotes the $i$th row in the $T^{-1}$ matrix.

Solving for $\nabla_{\bf x}\cdot T^{-1}$ in Eq.~(\ref{E:IS}) we
obtain:
\begin{equation} \label{E:ID2}
 {\nabla_{\bf x} \cdot T^{-1}
  = -\biggl(\sum_{i = 1}^{6}
   T^{-1}(i,:){\partial T\over \partial x_{i}}\biggr)T^{-1}}
\end{equation}

This simplification has also allowed us to express the divergence
independent of $\mathcal{H}$ once combined with Eq.~(\ref{E:MAT4})
as follows (this will prove to be useful later.)
\begin{equation}\label{E:18}
\begin{split}
 \nabla_{\bf x} \cdot \dot {\bf x}
  & = -\biggl(\sum_{i = 1}^{6} T^{-1}(i,:){\partial T\over \partial x_{i}}
   \biggr)T^{-1}\nabla_{x} \mathcal{H} ({\bf x}) \\
  & = -\biggl(\sum_{i = 1}^{6} T^{-1}(i,:)
   {\partial T\over \partial x_{i}}\biggr) \dot {\bf x}
\end{split}
\end{equation}

After fully expanding and condensing the operations in parentheses
above we obtain a useful and simplified expression:
\begin{subequations}\label{E:1}
\begin{align}
 \sum_{i = 1}^{6} T^{-1}(i,:){\partial T\over \partial x_{i}}
  & = {{e\over \hbar}\over (1 + {e \over \hbar}{\bf B} \cdot {\bf \Omega})}
  \begin{pmatrix}
  {\partial B_{2} \over \partial r_{1}}\Omega_{2}
    + {\partial B_{3} \over \partial r_{1}}\Omega_{3}
    - {\partial B_{2} \over \partial r_{2}}\Omega_{1}
    - {\partial B_{3} \over \partial r_{3}}\Omega_{1} \\
   {\partial B_{1} \over \partial r_{2}}\Omega_{1}
    + {\partial B_{3} \over \partial r_{2}}\Omega_{3}
    - {\partial B_{1} \over \partial r_{1}}\Omega_{2}
    - {\partial B_{3} \over \partial r_{3}}\Omega_{2} \\
  {\partial B_{1} \over \partial r_{3}}\Omega_{1}
    + {\partial B_{2} \over \partial r_{3}}\Omega_{2}
    - {\partial B_{1} \over \partial r_{1}}\Omega_{3}
    - {\partial B_{2} \over \partial r_{2}}\Omega_{3} \\
  {\partial \Omega_{2} \over \partial k_{1}}B_{2}
    + {\partial \Omega_{3} \over \partial k_{1}}B_{3}
    - {\partial \Omega_{2} \over \partial k_{2}}B_{1}
    - {\partial \Omega_{3} \over \partial k_{3}}B_{1} \\
  {\partial \Omega_{1} \over \partial k_{2}}B_{1}
    + {\partial \Omega_{3} \over \partial k_{2}}B_{3}
    - {\partial \Omega_{1} \over \partial k_{1}}B_{2}
    - {\partial \Omega_{3} \over \partial k_{3}}B_{2} \\
  {\partial \Omega_{1} \over \partial k_{3}}B_{1}
    + {\partial \Omega_{2} \over \partial k_{3}}B_{2}
    - {\partial \Omega_{1} \over \partial k_{1}}B_{3}
    - {\partial \Omega_{2} \over \partial k_{2}}B_{3}
  \end{pmatrix}^{\tau} \label{E:fir}   \\
  & =  {\nabla_{\bf x}(1 + {e\over \hbar}{\bf B} \cdot {\bf \Omega})
        \over (1 + {e\over \hbar}{\bf B} \cdot {\bf \Omega}) }
   \label{E:sec}
\end{align}
\end{subequations}

Eq.~(\ref{E:sec}) is revealed from the fact that the divergence of
the Berry curvature, ${\bf \Omega}$, as well as the magnetic field, ${\bf B}$,
are equal to zero.  Both are defined as the curl of a vector
function and the divergence of a curl must be zero, $\nabla \cdot
\nabla \times {\bf f} = 0$:
\begin{subequations} \label{E:DivBO}
\begin{align}
 \nabla_{\bf r} \cdot{\bf B}
  & = {\partial B_{1} \over \partial r_{1}}
      + {\partial B_{2} \over \partial r_{2}}
      + {\partial B_{3} \over \partial r_{3}} = 0 \label{first} \\
 \nabla_{\bf k} \cdot{\bf \Omega}
  & = {\partial \Omega_{1} \over \partial k_{1}}
    + {\partial \Omega_{2} \over \partial k_{2}}
    + {\partial \Omega_{3} \over \partial k_{3}} = 0 \label{second}
\end{align}
\end{subequations}

In the first row of (\ref{E:fir}), for example, $-{\partial B_{2}
\over \partial r_{2}}\Omega_{1} - {\partial B_{3} \over \partial
r_{3}}\Omega_{1}$ is equal to simply ${\partial B_{1} \over
\partial r_{1}}\Omega_{1}$.  Using the solution in (\ref{E:sec})
our phase divergence, Eq.~(\ref{E:18}), reduces to a final
condensed form as follows.  Here $ J({\bf x}) = 1 + {e \over \hbar}
{\bf B}\cdot{\bf \Omega} $.
\begin{subequations}
\begin{align}
 \nabla_{{\bf x}}\cdot \dot{\bf x}
  & = - {\nabla_{\bf x} J({\bf x}) \over J({\bf x})}\dot {\bf x} \\
  & = - {d\over dt}\ln J({\bf x}) \\
  & = - {d \over dt} \ln\left(1 + {e \over \hbar}{\bf B} \cdot{\bf \Omega}\right)
     \label{E:divF}
\end{align}
\end{subequations}

We see that the divergence is time-varying given that both ${\bf
B}$ and ${\bf \Omega}$ are non-zero and non-orthogonal.  Thus our
system phase space is, in general, compressible, that is the
Liouville theorem does not hold. This feature was used in
Ref.\cite{xiao} as the indication that the dynamics were
non-Hamiltonian. However, different opinions have been voiced
\cite{bliokh,duval,ghosh,gosselin}. A related point will be
explicitly exposed in the next section.

\section{Jacobi Identity}

In this section we follow a standard procedure in mathematical
physics \cite{arnold,darboux} to determine when a system can be
(locally) regarded as a Hamiltonian system. In this regard, there
are two important conditions the system has to satisfy. The first
is the existence of a Poisson bracket which must be explicitly
anti-symmetric. The second is on the validity of the Darboux
theorem. The incompressible condition, or the Liouville theorem,
does not appear to be essential in this regard.

The existence of an antisymmetric Poisson bracket has already been
defined in the previous section, Eq.(\ref{poisson}), which endows
the system with a specific dynamical structure which may not be a
(local) canonical Hamiltonian system. For the Darboux theorem,
which guarantees that a non-degenerate system can be expressed in
local canonical coordinates, the crucial condition to it's
validation is the presence of the Jacobi identity.

We express the Jacobi identity as the sum of the cyclic
permutations of the double Poisson bracket.  Here $f(\eta)$,
$g({\eta})$, and $h(\eta)$ are any functions with continuous second
derivatives and $\eta$ is the system space in consideration.  The
Jacobi identity is:
\begin{equation}
   [f,[g,h]] + [h,[f,g]] + [g,[h,f]] = 0
\end{equation}

Direct evaluation to verify the Jacobi identity would involve a
great deal of algebra. By computing the Poisson bracket as a
summation over each vector field we notice most of the terms
cancel out.
\begin{equation}
 [f,g] = \sum_{i,j = 0}^{n}{\partial f \over \partial
    x_{i}}\phantom{x} T^{-1}(i,j)\phantom{x} {\partial g
     \over \partial x_{j}} \, .
\end{equation}
Here $T^{-1}(i,j) = Q_{ij}({\bf x}) $ is the $ij$th element of the
$T^{-1}$ matrix. Also to more easily display derivatives a
notation will be used in which ${\partial f / \partial x_{i}}$ is
simply $f_{i}$. Evaluating the first double Poisson bracket
produces,
\begin{equation}
\begin{split}
 [f,[g,h]]
  & = \sum_{i,j,k,l =0}^{n} f_{i}T^{-1}(i,j) \Bigl(g_{k}T^{-1}(k,l)h_{l}\Bigr)_{j}\\
  & = \sum_{i,j,k,l =0}^{n} f_{i}T^{-1}(i,j) \Bigl( g_{k,j}T^{-1}(k,l)h_{l}
      + g_{k}T^{-1}(k,l)h_{l,j} +g_{k}T^{-1}_{j}(k,l)h_{l}\Bigr)
\end{split}
\end{equation}

The next two double Poisson brackets give similar results and
because $T^{-1}$ is antisymmetric the first 2 terms in the
parentheses of each will cancel out \cite{goldstein}.  In the
simple case in which the transformation matrix is antisymmetric
and independent of $\bf x$ this will alone satisfy the identity.
However, difficulty arises because ${dT^{-1} / dx_{i}} \neq 0$ and
we are left with the third term of each permutation containing the
derivative of the $T^{-1}$ matrix as follows:
\begin{equation}
 [f,[g,h]] + [h,[f,g]] + [g,[h,f]]
  = \sum_{i,j,k,l = 0}^{n} f_{i}g_{k}h_{l}\Bigl(T^{-1}(i,j)T^{-1}_{j}(k,l)
    + T^{-1}(l,j)T^{-1}_{j}(i,k)+ T^{-1}(k,j)T^{-1}_{j}(l,i)\Bigr)
\end{equation}

Eq.~(\ref{E:JF}) is arranged to collect the arbitrary $f,g,h$
functions.  Because each $f_{i}g_{k}h_{l}$ is unique the
expression in parentheses must equal zero for every combination of
$i,k,l$ in order to satisfy the identity.
\begin{equation} \label{E:JF}
  \sum_{j = 0}^{6} T^{-1}(i,j)T^{-1}_{j}(k,l)
     + T^{-1}(l,j)T^{-1}_{j}(i,k)+ T^{-1}(k,j)T^{-1}_{j}(l,i) \quad
   = 0 \qquad \text{for all $i,k,l$}
\end{equation}

Because $i,k,l = 1,2,...,6$ there are $6^{3}$ equations that we
must prove are equal to zero. However, we have been able to express all
of them in four general forms through index notation.  To show how
this is possible it is helpful to first state the $T^{-1}$ matrix in
index notation.  Here each quadrant is a $3\times3$ matrix with
$a$ denoting row and $b$ column
\begin{equation} \label{E:index}
 {T^{-1}({\bf x})
  = {1\over \sqrt{\det(T)}}\begin{pmatrix}
    \varepsilon_{abc} \Omega_{c}
     & \delta_{ab}
    + {e \over \hbar}B_{a} \Omega_{b} \cr
    - \delta_{ab} - {e \over \hbar}\Omega_{a} B_{b}
     & -\varepsilon_{abc} {e \over \hbar}B_{c}
                           \end{pmatrix}} \; ,
\end{equation}
with the determinant
\begin{equation}
 \det(T({\bf x})) = \left(1 + {e\over\hbar}{\bf B} \cdot{\bf \Omega} \right)^{2} \, .
\end{equation}

When solving Eq.~(\ref{E:JF}) with the index notation given in
(\ref{E:index}) a distinct difference in solutions is noticed when
$i,k,l$ have values of $1,2,$ or $3$ and when they have values of
$4,5,$ or $6$.  To take advantage of this we will now improve our
notation of each subset of $i,k,l$.  A value between $1$ and $3$
will be denoted by $r$ and between $4$ and $6$ by $k$. Particular
values in $r$ and $k$ can now be described by integers between 1
and 3 for the sake of index notation.  In the following solutions
these integers are given, again, by $a,b,$ and $c$ which are
unrelated to those in (\ref{E:index}).  The specific process to
arrive at each solution is straightforward but lengthy and is
excluded here.  The final simplified solution for each subset is
as follows.
\begin{subequations}
\begin{align}
 (r_{a}, r_{b}, r_{c})
  & =  {\varepsilon_{abc} \over \det(T)}(\nabla_{k}\cdot{\bf \Omega}) \\
 (r_{a}, k_{b}, k_{c})
  & = {\varepsilon_{bcd}({e\over\hbar})^{2}B_{a}B_{d}\over \det(T)}(\nabla_{k}\cdot{\bf \Omega})
     + \delta_{ab}{{e\over\hbar}\Omega_{c}\over \det(T)}(\nabla_{r}\cdot {\bf B})
     - \delta_{ac}{{e\over\hbar}\Omega_{b}\over \det(T)}(\nabla_{r}\cdot {\bf B})  \\
 (k_{a}, k_{b}, k_{c})
  & = {\varepsilon_{abc}{e\over\hbar} \over \det(T)}(\nabla_{r}\cdot {\bf B}) \\
 (k_{a}, r_{b}, r_{c})
  & = {\varepsilon_{bcd}{e\over\hbar} \Omega_{a}\Omega_{d}\over \det(T)}(\nabla_{r}\cdot {\bf B})
     + \delta_{ab}{{e\over\hbar}B_{c}\over \det(T)}(\nabla_{k}\cdot {\bf \Omega})
     - \delta_{ac}{{e\over\hbar}B_{b}\over \det(T)}(\nabla_{k}\cdot {\bf \Omega})
\end{align}
\end{subequations}

As previously mentioned in Eq.~(\ref{E:DivBO}) the divergence of
both Berry curvature in momentum space and the magnetic field in
position are zero and each subset above will also be zero.  It
should also be noticed that the above equations only accommodate
one half of the $6^{3}$ total equations. The rest come from
permutations in $(r,k,k)$ and $(k,r,r)$ and are similar
expressions.

Thus the Jacobi identity is satisfied for the Bloch-Peierls-Berry
dynamics. Hence the Darboux theorem holds. Therefore, these
dynamics can be locally mapped into a canonical form of Hamiltonian
dynamics, and may be named Hamiltonian dynamics. In fact
a global transformation has even been suggested \cite{bliokh}. The
reformulation of Bloch-Peierls-Berry dynamics into Darwinian
dynamics makes this point evident from a completely different
perspective.

%

\section{ Fokker-Planck Equation and Equilibrium Distribution }

The general form of the present Darwinian dynamics described by
Eq.~(\ref{E:Darwin}) has been shown to correspond to a
Fokker-Planck equation describing probability evolution in phase
space \cite{ao2005,kat,ya2006}:
\begin{equation} \label{distribution}
 {\partial \rho({\bf x}, t)  \over \partial t}
  = \nabla_{\bf x}^{\tau} \cdot \Bigl [
   \bigl[D({\bf x}) + Q({\bf  x}) \bigr]
    \bigl[ \omega  \nabla_{\bf x}
     + \nabla_{\bf x}\Phi({\bf x}) \bigl] \Bigr] \rho({\bf x}, t)
\end{equation}
Here $\omega$ is a non-negative constant equivalent to
temperature. $D$ is a symmetric matrix and $Q$ is an antisymmetric
matrix so that
$$
  (D + Q) = (S + T)^{-1} .
$$

Applying Eq.(\ref{distribution}) to the Bloch-Peierls-Berry
equation as given in Eq.~(\ref{E:dx}) we find that $D = 0$ and $Q
= T^{-1}$, thus there is only an anti-symmetric part and the
diffusion matrix $D$ is in the zero limit:
\begin{equation} \label{E:FP}
 {\partial \rho({\bf x}, t) \over \partial t }
  = \nabla_{\bf x}^{\tau} \cdot \Bigl[ Q({\bf  x})
      \bigl[ \omega  \nabla_{\bf x}
     + \nabla_{\bf x}\Phi({\bf x}) \bigl] \Bigr] \rho({\bf x}, t)
\end{equation}
The term on the far right side of the equation can be seen to
represent the probability flow divergence, $\nabla_{\bf x}\cdot\bigl(
T^{-1}\nabla_{\bf x}\Phi\rho\bigr) \Rightarrow
\nabla_{\bf x}\!\cdot\!(\dot {\bf x}\rho)$, by substituting
Eq.~(\ref{E:dx}).  Thus in the trivial case where $\omega = 0$
Eq.~(\ref{E:FP}) becomes the standard form of the continuity
equation.  The term $\nabla_{\bf x}\cdot( \omega \nabla_{\bf x} \cdot
T^{-1})\rho$ in Eq.~(\ref{E:FP}) represents the additional effect
on distribution at non-zero temperatures.

We now provide a plausible argument for another choice of
equilibrium distribution other than that in Ref.\cite{xiao}.
First, we note that in the anti-symmetric matrix $T$ in
Eq.(\ref{E:dx}), the geometric phases due to the magnetic field
${\bf B}$ \cite{berry,geometricphase} and the novel Berry
curvature ${\bf \Omega}$ are in equal footing. They all contribute
in the same manner to the geometric phase for a possible close
trajectory in the phase space. A unifying description of the
dynamics is hence achieved in this Darwinian dynamics formulation.

Second, in the case that dynamics are dissipative in phase space,
for example if there are indeed electron-phonon and
electron-impurity interactions, then the friction or resistance matrix
$S$, and hence the diffusion matrix $D$, is not zero. Such a
dissipative dynamics is very likely to be described by
Eq.(\ref{E:Darwin}). Further, such a dissipative dynamics in the
presence of Berry phase has been well characterized in condensed
matter physics, such as the dynamics of topological singularities
in superconductors and superfluids \cite{at}. Thus, extending
Bloch-Peierls-Berry dynamics to include dissipation as in the form
of Eq.(\ref{E:dx}) or (\ref{E:Darwin}) indeed appears to be a
natural choice.

Based on the above considerations we may adopt Eq.(\ref{distribution})
as the equation for dynamics of non-interacting electrons in the
presence of dissipation. The equilibrium distribution for such
dynamics immediately reads:
\begin{equation} \label{equilibrium}
 \rho_{eq}({\bf x}) = \frac{1}{\cal Z}
   \exp\left\{ - \frac{ \Phi({\bf x}) }{\omega} \right\} \; .
\end{equation}
The partition function summing up all probability in phase
space may be defined as
\begin{equation} \label{equilibrium}
  {\cal Z} = \int d^6{\bf x} \;
   \exp\left\{ - \frac{ \Phi({\bf x}) }{\omega } \right\} \; .
\end{equation}
The time dependent free energy for such an open system may be
defined as
\begin{equation} \label{equilibrium}
 {\cal F}(t) = - \omega \int d^6{\bf x} \; \rho({\bf x}, t)
  \ln\left(\frac{\rho({\bf x}, t)}{\rho_{eq}({\bf x})}\right) + {\cal F}_{eq} \; ,
\end{equation}
which always decreases towards the equilibrium value $  {\cal
F}_{eq} = - \omega \ln {\cal Z}$.

We note that if treating the equilibrium distribution $\rho_{eq}$
as a single particle distribution, one immediately notices a
difference between the present result and the one obtained
in Ref.\cite{xiao} ({\it c.f.} their reference [21]) in the same
limit.

\section{Discussions}

We have investigated the Bloch-Peierls-Berry dynamics from a
nonequilibrium dynamical point of view. The Bloch-Peierls-Berry
equations can be reformulated into a simple and generic form. Using
this reformulation we have explicitly demonstrated the compressibility
of Bloch-Peierls-Berry dynamics and the embedded Jacobi identity,
both were pointed out previously in literature. From the point of
view of Hamiltonian dynamics, we explicitly showed that the
violation of the \lq\lq naive" Liouville theorem is not essential. At
the same time we reached a distribution function similar to what
is implied in Ref.\cite{xiao}.

There are two new features in our study that we would like to
point out. The first one is that our study is in the classical
domain, and is for one electron. No many-body effect, as strongly
suggested in Ref.\cite{xiao}, has been considered here. Hence,
from the condensed matter physics point of view, our work may be
relevant in the dilute and non-interacting limit, with a
relatively high temperature. In this sense, it would be further
interesting theoretically to see how the Boltzmann equation
implicitly discussed in Ref.\cite{xiao} would relate to the
present Fokker-Planck equation. In particular, how the
electron-phonon type interaction with explicit energy dissipation
would be incorporated into formulation in Ref.\cite{xiao}.

The second feature is that due to the open system nature our
probability dynamics formulation is necessarily in the domain of
the canonical ensemble, where the Boltzmann-Gibbs distribution is
emphasized. This also implies that there is a preferred and
natural choice of phase space. Instead, the discussion of
Hamiltonian flow in Ref's.\cite{bliokh,duval} appears within the
micro-canonical ensemble. It is well known that the transition
from micro-ensemble to canonical ensemble is not unique. This may
be the reason that within the present nonequilibrium formulation
there is no compelling reason to emphasize the Liouville theorem,
even though the Hamiltonian structure is evident in both
approaches.

In conclusion, the present exploration from the Darwinian dynamics
perspective has further revealed the richness of
Bloch-Peierls-Berry dynamics.



{\ }

We thank D. Xiao and Q. Niu for bringing
Ref's.~\cite{bliokh,duval} to our attention.  This work was
partially supported by NIH grant \#HG002894.


\end{document}